\documentclass[10pt,conference]{IEEEtran}

\usepackage{graphicx}

\usepackage{amsmath}
\usepackage{framed}
\usepackage[tight,footnotesize]{subfigure}
\usepackage[geometry]{ifsym}
\usepackage{algorithmic}
\usepackage{algorithm}
\usepackage{amssymb}
\usepackage{url}
\usepackage{cite}

\usepackage{paralist}
\usepackage{enumerate}

\begin{document}

\title{Object-Oriented Networking}
\author{
\IEEEauthorblockN{Panos Georgatsos}
\IEEEauthorblockA{CERTH,\\
Greece.\\
pgeorgat@iti.gr}\and \IEEEauthorblockN{Paris Flegkas}
\IEEEauthorblockA{CERTH,\\
Greece.\\
pflegkas@uth.gr} \and \IEEEauthorblockN{Vasilis Sourlas}
\IEEEauthorblockA{UCL,\\
UK.\\
v.sourlas@ucl.ac.uk} \and \IEEEauthorblockN{Leandros Tassiulas}
\IEEEauthorblockA{Yale,\\USA.\\
leandros.tassiulas@yale.edu}}

\maketitle

\begin{abstract}
We propose the object-oriented networking (OON) framework, for meeting the generalized interconnection, mobility and technology integration requirements underlining the Internet. In OON, the various objects that need to be accessed through the Internet (content, smart things, services, people, etc.) are viewed as network layer resources, rather than as application layer resources as in the IP communications model. By abstracting them as computing objects -with attributes and methods- they are identified by expressive, discoverable names, while data are exchanged between them in the context of their methods, based on suitably defined system-specific names. An OON-enabled Internet is not only a global data delivery medium but also a universal object discovery and service development platform; service-level interactions can be realized through native network means, without requiring standardized protocols. OON can be realized through existing software-defined networking or network functions virtualization technologies and it can be deployed in an incremental fashion.
\end{abstract}


\section{Introduction}\label{intro}

The Internet is challenged not only by the multitude and diversity of interconnected objects (content items, sensors, controllers, services, people etc.) but also by the mobile and virtualized nature of their hosting environment. Services and applications are becoming increasingly demanding and dynamic in nature, requiring access to various sets of data, other services, controlled devices or mobile users, at Internet scale. Requirements for instant object publishing, open and secure access are immense. The advent of hardware virtualization and fast computing technologies boosts scalability and cost-effectiveness of Internet operations and services however it amplifies mobility requirements. Data, services and applications are highly virtualized, migrating within and across data centers of the same or different providers. The integration of virtual computing and networking technologies at global scale has become a crucial issue.

The IP communications model is not readily fit for meeting the above challenges. Communicating objects are viewed as application layer resources, accessed through specialized protocols, and networking is between network endpoints. This design makes it hard to follow object mobility and migration across the Internet, while it can lead to confined object availability within large-scale service providers, as we witness today; which, in our view, inhibits the provisioning of Internet-scale applications.

The evolution towards an all-connected world, the Internet of Everything (IoE) \cite{IOE}, which is underlined by strong mobility and migration requirements, requires a new set of global networking abstractions and related functions, beyond those of the IP model.

Network programmability and virtualization technologies, most notably SDN \cite{SDN,openflow} and NFV \cite{NFV}, are versatile means for building networking solutions, with openness and flexibility in incorporating network functionality being their main strengths; per se they do not constitute specific network architectures. Although already applied in specific environments, especially cloud environments, Internet architectures utilizing their benefits remain an active topic of research. 

The named networking nature of the ICN paradigm \cite{ccn}-\cite{icnsurvey} has the potential of gracefully meeting the instant availability and generalized mobility requirements of the IoE because it can address these requirements through native network means rather than through new/add-ons in protocols at the network and/or the application layers. The unique capability of ICN compared to host-centric networking for dealing with the location and service dynamics in an IoT environment \cite{IOT-survey} is highlighted in \cite{IOT-ICN}. Being content-centric, existing ICN architectures adopt a pull (request-reply) model for transferring data over the network. Their ability to support conversational and notification services is not evident. Although a number of proposals have been made (e.g., \cite{VoCCN,chilop}), especially on CCN/NDN architectures, service support in ICN is an open research issue.

We propose the \emph{object-oriented networking (OON)} paradigm to the end of providing a flexible and sustainable networking and service enablement infrastructure for the IoE. Following ICN, OON views the various objects that need to be accessed or communicate via the Internet as network layer resources identified by location-independent names. By abstracting them as computing objects, with attributes and methods, rather than as content (static or dynamic) like in existing ICN architectures, OON provides for:
\begin{itemize}
\item multiple-attribute descriptive object names, 
\item object discovery based on description semantics even with partially-specified names and
\item any form of data exchange (pull, push or interactive) between the methods that objects expose. 
\end{itemize}
By placing data exchange between objects in the context of their methods, required transport and application layer interactions may not necessarily rely on standardized protocols, while service development can be facilitated. Building on its semantic richness, the design of OON allows the flexible incorporation of security-by-design solutions.

The rest of the paper is organized as follows. Section~\ref{principles}, \ref{data-layer} and~\ref{information-layer} describe the OON framework. Section~\ref{routing} discusses routing scalability in OON, while Section~\ref{realisation} addresses realization and deployment aspects. Section~\ref{benefits} presents the OON benefits, whereas Section~\ref{icn} positions OON with respect to existing ICN architectures. Finally, Section~\ref{conclusions} concludes the paper and highlights dimensions for future work.

\section{OON Notions and Principles} \label{principles}

\subsection{Physical and Informational Object Forms}\label{forms}

OON proposes a ``named object'' networking model, where the term ``object'' refers to anything that contains data e.g., documents, books, articles, pictures, videos and movies, or is able to produce or consume data e.g., smart meters, sensors, various services, users and individual persons. The various data-consuming, -producing or -holding objects are seen as instances of specific well-defined classes since they can be distinguished into (sub-)types, as indicated by the previous examples, that can be clearly and comprehensibly described. This premise is in line with various on-going modeling activities like the Dublin Core Initiative \cite{dublin} and the FOAF project \cite{foaf}.

As a result, objects can be viewed that exist in a class-instantiated or \emph{informational form}, comprising a set of attribute-value pairs and a list of methods, as appropriate to the class they belong. In addition, objects are viewed in their ``default'' \emph{physical form} in which they actually exist in a networked environment e.g., as a file, a computer process, a physical thing or a human being.

The following point is worth-making. Existing networking technologies are solely concerned with objects in their physical form, irrespective of whether they assume network location-specific or informational addressing schemes. In OON, we complement object physical forms with informational forms for abstracting networking at a user comprehensible level, which is beneficial from many aspects; object naming, discovery, higher-level interactions and service development.

An object may exist in multiple physical forms e.g., cached content items. However, it can only be associated with a single informational form, which signifies its availability -objects cannot be accessed unless their informational forms exist.

The attributes in the object informational forms are distinguished into: description attributes, presenting the main characteristics and properties of the object class; management attributes, holding the state, status and various use and usage statistics; and, relationship attributes, holding ``pointers'' to (the identifiers of) the corresponding object's physical form(s), perhaps other objects according to well-defined relationships.

The methods in the object informational forms correspond to the rudimentary actions that can be performed on the data that the object may hold, produce or consume, such as to push/send, pull/get or sink/consume data. For instance, a content item presents a method for sending its data. Additional methods may be provided to expose the particular higher-level communication services that a specific object class may offer e.g., to chat or to talk for persons or to pause or to jump for videos. It is worth noting that security and management functionality may be embodied in the object methods and/or in other specialized, not publicly visible though, objects. Without loss of generality (see Section~\ref{realisation}), it is assumed that the realization of the object methods is part of their physical forms.

\subsection{Data and Information Networking Layers}\label{layers}

OON encompasses two distinct networking layers: the \emph{information networking layer} and the \emph{data networking layer}. The data networking layer maintains and interconnects objects in their physical form and it is responsible for transferring data between objects. The information networking layer maintains and interconnects objects in their informational form and enables the discovery of objects that can be accessed at global scale.

It is noted that object discovery is based on the semantics of the attributes included in the object informational forms, not on the data (terms/key words) that the objects actually contain or produce. As such, OON does not obsolete the role of search engines. On the contrary, it facilitates required crawling operations, avoiding tensions with NSPs, and enhances the ability to search for everything as it becomes available in the Internet, including things and their data which currently are out of searching scope; these can be achieved by crawling first the information networking layer. Furthermore, OON calls for a distributed multi-polar search paradigm to the benefit of scalable global and instant search of everything.

Logically, the information networking layer lies above the data networking layer since object informational forms hold pointers to (the identifiers of) the corresponding object physical forms. The two layers can operate completely independently with the correctness of their operations depending on the consistency between the object forms in each layer. Appropriate mechanisms should therefore exist to ensure that the objects' informational forms maintain valid and up-to-date pointers to their physical forms.

The data and information networking layers are operated by different interconnected OON domains, which may not necessarily be in $1:1$ correspondence. Domains in different layers need not be connected, since each layer has distinct networking goals. As in today's Internet, the global topology at each networking layer cannot be known.

Domains at the data networking layer are mainly characterized by their geographical span, while domains at the information networking layer are mainly characterized by the volume and kinds of objects that they hold. In analogy to IP, domains at the data networking layer correspond to NSPs, while domains at the information networking layer correspond to those running the DNS backbone. In fact, the information networking layer can be viewed as a multiple-attribute naming resolution mechanism with inherent searching capabilities as opposed to a fixed-naming resolution service, as in IP.

\subsection{Object Naming}\label{naming}

Objects are identified in their physical and informational forms by distinct identifiers: a \emph{an informational name (i-name)} and a \emph{physical-form name (p-name)}, respectively.

Informational names identify objects in a descriptive manner at a user-friendly abstraction level. Each object class can be completely characterized in terms of the so-called \emph{class-defining attributes} -with objects within the class differing in their values- which by definition are part of the description attributes included in the object informational forms. As such, object i-names and informational forms are not different structures; i-names are included in informational forms and conversely, informational forms extend i-names with additional information. Objects are identified by their i-name in the information networking layer.

Physical-form names identify objects from a data networking perspective. The routing functions (route dissemination, aggregation, selection) in the data networking layer operate on these names. It is noted that for routing scalability and forwarding efficiency, objects could not be identified by their i-name, since i-names do not provide for high degrees of aggregation, while their structure (components and size) are largely variable. In addition to being hierarchical and of fixed length for enabling scalable and efficient routing, object p-names should not bear network location or service technology semantics for supporting object mobility and migration, while they should be consistent for facilitating caching and multicasting within and across domains. Note that bindings for verifying object integrity and provenance could be embedded in the object informational forms, not in p-names.

The definition of the semantics and the structure of object p-names should be seen in conjunction to their assignment process (see next section) and their ability to support scalable network operations. It is left as a design choice of different OON approaches, which of course are required to inter-operate at the data networking layer.

Since object i- and p-names serve distinct purposes -i-names are used for object discovery, while p-names are used for routing data between objects- they have different characteristics. Object i-names are expressive and comprehensible, being made up of common-sense attributes, which may even be guessed if not known, whereas p-names are system-centric and generally not user-friendly. Object i-names may be supplied in a partially or loosely defined form, whereas p-names are strict in syntax and value being of no use unless they are supplied correctly in their entirety. In analogy to IP networking, object i-names correspond to URIs and p-names to IP addresses.

\subsection{Overall OON Operation}\label{operation}

First, objects need to be \emph{instantiated} i.e., have their physical form appeared in data networking domain(s) and \emph{published} i.e., have their informational form created in an information networking domain. In the general case, object instantiation and publishing takes place at the same epoch, following a bottom-up, first instantiated then published, or a top-down, first published then instantiated, procedure. However, publishing may not be required for objects whose data networking name, p-name, can be made known to other objects through other specialized means than public discovery.

Objects have their informational name, i-name, defined, during publication, as part of the process of filling out their informational forms. This process may be carried out through automated and/or manual means, depending on particular (publishing and/or published) object characteristics.

Objects may be assigned their p-name by the domain holding their physical form, during instantiation, or by the overlying domain maintaining their informational form, during publishing. Evidently, this choice is of significant importance since it affects the semantics, structure and number of distinct prefixes of object p-names, therefore data routing scalability and forwarding efficiency, as well as name consistency across data networking domains. Since it touches upon business aspects, it is left as an option. In any case, for data routing scalability and name consistency reasons, OON assumes that there is a trusted organization like IANA \cite{iana}, which supplies domains (either in the information or in the data networking layer) with ``top level identifiers'' for \emph{prefixing (the physical forms of) objects in a systematic manner}, that is in a manner that favors aggregation per individual domain.

\begin{figure*}[!t]
\begin{center}
\includegraphics[width=5in]{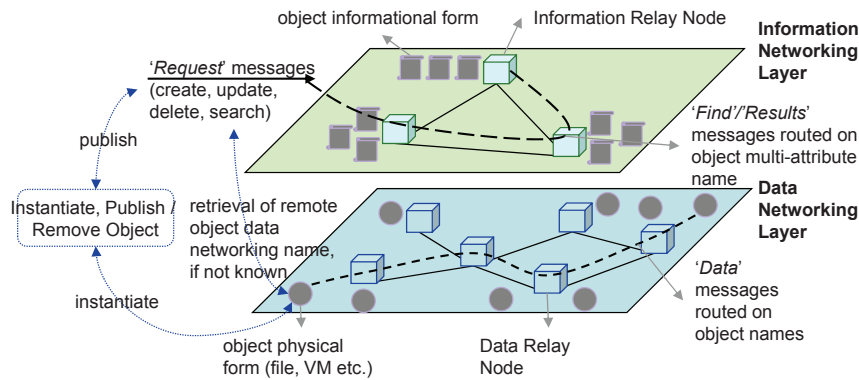}
\caption{OON model.} \label{architecture} 
\end{center}
\end{figure*}

Once objects are instantiated, they can exchange data with other objects, through the data networking layer, provided that their p-names are known to each other (through discovery or other means). Once objects are published, they can be discovered by other objects based on the semantics of their attributes, through interactions with the information networking layer. Their p-name can be retrieved and subsequently, if desired, data exchange can take place in the data networking layer. Figure \ref{architecture} presents the logical architecture of OON summarizing the section.

\section{Data Networking Layer}\label{data-layer}

The data networking layer provides for a single network message, the ``Data'' message, for carrying data between objects in their physical form.

\emph{Data exchange between objects is always done within the context of their methods}. The ``Data'' message header includes information about the calling object and its initiating method, the target object and the method to be invoked therein. This quadruplet is analogous to the source and destination pairs of IP address and port number included in the IP header. It also includes a reply-to method specified by the calling object, to which the called object should send subsequent data. The default for the calling, called and reply-to methods are the generic data-related methods -to get, send and sink data- that every object supports. Additional information may be included such as transfer priority and cumulated transfer time, as deemed necessary by specific OON approaches.

Generally speaking, objects can learn about the methods that other objects support by discovering the objects and ``reading'' their informational forms. The latter can be done manually or though automated inspection means. Note that the methods of widely used object classes may be globally known.

In the following we exemplify the ``Data'' header information through typical data exchanges. Note that object methods are named as actions that can be performed to the objects than as capabilities offered to other objects. That is, a content item exposes a ``send data'' method than a ``get data'' method for enabling other objects to retrieve its data.

If a data consumer, object $A$, wishes to retrieve the data of a file, object $B$, object $A$, say its method \texttt{GetDataFrom}, will issue a ``Data'' message to object $B$, calling for its method, say \texttt{SendDataTo}; subsequently, for sending the data to the requester, object's $B$ called method will issue ``Data'' messages to object $A$, calling for its method, say \texttt{SinkDataFrom}, or the particular method that object $A$ specified in the initial ``Data'' message.

The voice conversation between two persons, objects $A$ and $B$, involves an exchange of ``Data'' messages in both directions. Assuming that the two objects have agreed to converse, object $A$, say its method \texttt{Talking}, sends ``Data'' messages to object $B$, say to method \texttt{Listening}, specifying also that the reply data should be sent to its method, say \texttt{Listening}; and, similarly, in the other direction, object $B$ sends its own ``Data'' messages to object $A$.

``Data'' messages are routed on called-object information i.e., on the names identifying object physical forms (p-names, see Section \ref{naming}). Calling-object information or other information could be utilized for providing differentiated routing. It is taken that object methods exist within local object scope, implying that object physical forms cannot be distributed. As such, object method information cannot be globally identifiable and therefore used for routing.

The design choice of including not only object but also method information in the data-carrying messages is made to the benefits of security and higher-level communications. From a security perspective, it adds extra levers. Method names, in particular the reply-to method, could be cryptographic and/or could vary during sessions for continuously asserting that data are exchanged between legitimate parties. Higher-level communications are facilitated in terms of flexibility, performance and service provisioning as there is no need to standardize required protocols. Higher-level services can be considered as objects per se and therefore their interfaces -object methods- can be invoked natively by the proposed network message. Consensus on required capabilities and interface semantics can be established significantly faster than agreements on technical specifications through explicit standardization cycles.

Scalable means for inter/intra-domain routing based on physical-object names, associated protocols, mechanisms and generic nodal functions are the main challenges in this layer (see Section \ref{pnamer}). Caching and multicast, which are obviously facilitated by the named-object nature of OON, are not considered at a framework level; they are left as open design issues to be dealt within the context of specific OON approaches.

\section{Information Networking Layer}\label{information-layer}

The information networking layer provides for the ``xFind'' and the ``Results'' messages, which, in the general case, are in $1-N$ correspondence. Additionally, it provides for interface messages corresponding to the actions and their responses that may be performed on object informational forms, namely: find the ones matching certain criteria, register a new one, modify and delete existing ones. Interface action messages translate to appropriate ``xFind'' messages and the produced ``Results'' messages to interface responses.

In its find-version, the ``xFind'' message includes information about the object(s) of a specific class to be sought for. This information corresponds to a partially defined object informational form. Specifically, it contains pairs of the class description attributes and logical expressions for their values. The message traverses the distributed environment maintaining the object informational forms to the end of locating those matching the included information. As they are located, they are packaged in ``Results'' messages and returned to the node that issued the ``xFind'' message following the reverse route.

In its other versions, the ``xFind'' message includes the exact attribute-value pairs of the object's informational form to be created, modified or deleted. In these cases, the ``xFind'' message effectively checks the absence or existence of the carried informational form, which only if so is created or modified/deleted, respectively. A single ``Results'' message is returned containing an affirmation of the requested action.

Generic means for scalable and efficient routing of ``xFind'' messages in a distributed multi-domain environment that is, for routing on multiple-attribute names with precisely or loosely specified values, is the main challenge in this layer (see Section \ref{inamer}). In addition, procedures and mechanisms for aligning, merging and distributing object informational forms across domains are required. Obviously, these procedures are performed off-line at the granularity of agreement establishment between domains.

Mechanisms for secure object access should be incorporated. A policy-based paradigm is recommended, whereby access control policies could be defined by the objects themselves setting the viewing (of informational forms) and communication (data exchange) rights of requesting objects. The interface messages should convey an unambiguous identifier of the requesting object e.g., its informational form and be time-stamped.

This layer bears its own security concerns, which are related to its operational integrity and the validity/accuracy of the maintained information. Appropriate mechanisms should be provided for ensuring that the infrastructure, interfaces and information are not compromised; such as, mechanisms for detecting false object informational forms and for avoiding malicious attacks. These security concerns are an aftermath of OON, however they should be weighed out with its benefits; after all, they are common security concerns for which solutions have already been worked out.

The above security aspects, at object access and layer operations levels call for a logically distinct \emph{information security layer}, where all related mechanisms could be realized. The architecture of such a security layer is orthogonal to the proposed OON framework.

\section{Routing in OON} \label{routing}

\subsection{Routing on object i-names}\label{inamer}

For scalable and efficient routing based on multiple-attribute names, we propose a \emph{lexicographic partitioning-based routing} scheme, ala multiple-dimension DHT. Such names can be seen as points in a multi-dimensional information space, with the dimensions corresponding to the attributes. The value space in each attribute-dimension is ordered and bound (e.g., A*-*Z) and as such it can be segmented. By taking unions of the Cartesian product of these elementary segments, n-cubes, lexicographic partitions can be formed, which are then assigned to suitably connected nodes. This gives rise to a network of \emph{Information Relay Nodes (IRNs)}, each holding and managing a specific subset of the whole multiple-attribute namespace.
 
The proposed lexicographical approach to partitioning and networking a multiple-attribute namespace avoids the requirement for an explicit exchange of routing information as objects are added/updated/deleted and facilitates simple forwarding schemes. As the namespace is partitioned and pre-assigned to IRNs, at publishing epochs, (the informational form of) an object just needs to be forwarded to the node assigned to maintain the corresponding partition. Similarly, at request epochs, that node needs to be reached. The formation of lexicographic partitions and the topology of the IRN network are the free parameters for tuning to best meet the intrinsic aspects of routing in multi-dimensional namespaces. We have worked (to be published) in such schemes and the results confirm scalable and efficient operations at the expense of computing and connectivity resources. 

\subsection{Routing on object p-names}\label{pnamer}

OON prescribes (see Section \ref{naming}) that object p-names are hierarchical with fixed number of components and size. Without loss of generality it can be assumed that their structure if of the form $\left\langle \text{GlobalId/LocalId}\right\rangle$, where the first part should be globally unique and is used for inter-domain routing, while the second part should be unique within the first part and is used for intra-domain routing. As outlined in Section \ref{operation}, object p-names are assigned by providers holding the object informational or physical forms. As such, the magnitude of $\left\langle \text{GlobalId}\right\rangle$ is in the order of the number of such providers, not of the objects, which obviously impacts positively on inter-domain routing scalability. Routing on p-names can follow the routing schemes proposed in the literature for ICN routing.  

\section{Realization and Deployment Aspects} \label{realisation}

The data and information networking layers can be realized through existing software-defined networking or network functions virtualization technologies. They can be deployed in an incremental fashion. The information networking layer can initially be deployed to provide for object discovery and resolution to IP addresses or URIs. Subsequently, the data networking layer can be deployed, again in an  incremental fashion e.g., for different types of objects. It is noted that the realization technologies facilitate deployment and interoperation with IP.

As stated in Section \ref{forms}, the realization of the objects' methods is considered to be part of their physical form. Evidently, this is the case for service kind of objects. For other kinds of objects, this can be achieved through alternative means. Indicatively, we mention: by specialized ``container objects'' or by transparent applications e.g., OON servers, in which cases they should appear in the OON object name space on behalf of the objects that they hide; by native means at file system or OS levels, in the end-system where the object physical forms exist.

The following points are worth-making. The notion of containers is not only useful for deployment but also for scalability reasons. Transport control logic could be provided as part of the realization of object methods or through specialized objects, which evidently should not be publicly accessible. Similarly, network and service management intelligence could be provided in the form of OON-adhering objects, without requiring standardized protocols for their interactions. Overall, OON proposes an \emph{object-oriented communications model} advocating an open communications
software market.

\section{OON Benefits} \label{benefits}

By design, OON provides for seamless data transfer and semantic discovery of objects across an Internet-connected world (of content, things, services and people) through common means, \emph{avoiding interoperability problems between different naming and networking systems for different kinds of objects}. Building on the ICN paradigm (see Section~\ref{principles}), its ``named-object'' networking nature can \emph{inherently support mobility, migration, in-network caching and multicast}.

By distinguishing between object informational and physical forms, OON allows the \emph{flexible accommodation of security-by-design solutions} for object verification and access control, identity management and privacy, without burdening the naming structure used for data networking.

OON promotes a communications model, where higher level interactions may not \emph{necessarily be based on standardized protocols}, but rather on a cascade of method calls; in OON, service capabilities can be exposed in a technology-agnostic manner, may be discovered if not a priori known and invoked by native network means. This obviously impacts positively on Internet service provisioning.

Finally, OON constitutes a \emph{sustainable networking and service enablement infrastructure} not only for the evolving dynamics of the current Internet but also for future requirements. Required functionality, being basically software, can be introduced in the Internet as needed in the form of OON-adhering objects.

\section{OON and ICN} \label{icn}

OON follows the named-networking principle of ICN however it does not build around the notion of content as existing ICN architectures do. To us, the ICN paradigm is orthogonal to content. By bringing OOP principles into ICN, OON builds around the notion of named object (not just named data)- in the OOP sense, an object comprises attributes (data and content per se) and methods through which data can be accessed. As such, OON differs from existing ICN architectures and brings unique benefits over them:
\begin{itemize}
\item Semantically rich names; in current ICN architectures names are hierarchical or flat, not of a descriptive multiple-attribute structure as in OON that evidently bears richer semantics.
\item Discovery and dynamic name construction capabilities; in current ICN architectures names should be supplied as provided by the producer/publisher, whereas in OON they may be supplied in a partially-specified form, being filled in automatically and selected by the user (request object) if multiple options exist.
\item Native means for bridging user and network views of the world - because of the above. Note that this does not obsolete current Web search engines, which provide for deep-data-(content)-based search, whereas OON can only provide for meta-data-based search.
\item Native support of any service interaction model; current ICN architectures are receiver-initiated (providing for a pair of data retrieval messages, request and reply messages), whereas OON can natively support data exchange between everything in any possible mode -pull, push or interactive.
\item Native support for higher-level services; current ICN architectures may require dedicated standardized protocols at transport and application layers, whereas OON advocates a protocol-less communication model - required intelligence (brought as objects) and interactions (realized through method-oriented data exchange) can be deployed faster than waiting standardization.
\end{itemize}
Generalizing ICN from named-data to named-object networking, all existing ICN architectures can be mapped to OON. For example, the CCN/NDN architecture \cite{ccn} maps to the data-networking layer of OON with the Interest and Data messages corresponding to the OON Data messages targeted for send-data and sink-data methods, respectively. The upper layer of OON, the information-network layer, that brings the three first of the above benefits, is missing from all ICN architectures.               

\section{Conclusions and Future Work} \label{conclusions}

We proposed object-oriented networking (OON) as an Internet architectural framework, for meeting current and future interconnection, mobility, migration and technology integration requirements. The key element of the OON design is the interconnection of the various entities that communicate or are accessed via the Internet from an informational perspective.This enables distributed, multi-polar search of everything based on description semantics and the realization of higher-level interactions by network-native means. Aspects of future work correspond to the development and performance evaluation of the technical challenges underlining the operation of the data and information networking layers: routing based on multiple-attribute names; schema of object informational forms and naming structure for data networking; procedures for object instantiation and publishing; TE and (self-)management functions and mechanisms at the information and data networking layers; and, security infrastructure and
mechanisms.



\begin{thebibliography}{01}

\bibitem{IOE} D. Evans, ``The Internet of Everything: How More Relevant and Valuable Connections Will Change the
World'', 2012. https://www.cisco.com/web/about/ac79/docs/ innov/IoE.pdf

\bibitem{SDN} ``Software-Defined Networking: The New Norm for Networks'' White paper - Open Networking Foundation, 2012.

\bibitem{openflow} N. McKeown, T. Anderson, H. Balakrishnan, G. Parulkar, L. Peterson, J. Rexford, S. Shenker and J.
Turner, ``OpenFlow: Enabling innovation in campus networks,'' ACM SIGCOMM Computer Communication Review, vol. 38, April
2008.

\bibitem{NFV}  ``Network Functions Virtualisation -- Introductory White Paper'', SDN and OpenFlow World
Congress, Darmstadt-Germany. 2012. http://portal.etsi.org/nfv/nfv\_white\_paper.pdf

\bibitem{ccn} V. Jacobson, D. K. Smetters, J. D. Thornton, M. F. Plass, N. Briggs, R. Braynard, ``Networking named content,'' in ACM CoNEXT 2009.

\bibitem{pursuit} D. Trossen and G. Parisis, ``Designing and Realizing an Information-Centric Internet'', in IEEE Communications Magazine, vol. 50, issue 7, pp. 60-67, 2012.

\bibitem{comet} W. K. Chai, N. Wang, I. Psaras, G. Pavlou, Ch. Wang, G. G. de Blas, F. Salguero, L. Liang, S. Spirou, An. Beben and El. Hadjioannou,
``CURLING: Content-ubiquitous resolution and delivery infrastructure for next-generation services,'' in IEEE
Communications Magazine, vol. 49, pp. 112--120, 2011.

\bibitem{SAIL} B. Ohlman et al, ``First NetInf architecture description," April 2009. http://www.4ward-project.eu/index.php?s=file\_download\&id=39.

\bibitem{icnsurvey} G. Xylomenos, C. N. Ververidis, V. A. Siris, N. Fotiou, C. Tsilopoulos,
X. Vasilakos, K. V. Katsaros and G. C. Polyzos, ``A Survey of Information-Centric Networking Research,'' in IEEE
Commun. Surveys, 2013.

\bibitem{IOT-survey} L. Atzori, A. Iera and G. Morabito, ``The Internet of Things: A survey'', Computer Networks - Elsevier, vol. 54, pp. 2787--2805, 2010.

\bibitem{IOT-ICN} Y. Zhang, D. Raychadhuri, R. Ravindran and G. Wang, ``ICN based Architecture for IoT'', IETF Internet Draft, ICNRG WG, Dec. 2013.

\bibitem{VoCCN} V. Jacobson, D. K. Smetters, N. H. Briggs,  M. F. Plass, P. Stewart, J. D. Thornton and R. L. Braynard, ``VoCCN: Voice-over Content-centric Networks'', Workshop on Re-architecting the Internet (ReArch), pp. 1--6, 2009.

\bibitem{chilop} C. Tsilopoulos and G. Xylomenos, ``Supporting Diverse Traffic Types in Information-centric Networks,'' ACM SIGCOMM ICN workshop, 2011.

\bibitem{dublin} Dublin Core Metadata Initiative (DCMI), http://dublincore.org/

\bibitem{foaf} The Friend of a Friend project (FOAF), http://www.foaf-project.org/

\bibitem{iana} Internet Assigned Numbers Authority. https://www.iana.org/

\end{thebibliography}
\end{document}